\documentclass[a4paper,fleqn,usenatbib]{mnras}

\usepackage[T1]{fontenc}
\usepackage{ae,aecompl}

\usepackage{graphicx}	
\usepackage{amsmath}	
\usepackage{amssymb}	

\newcommand{\be}{\begin{equation}}
\newcommand{\ee}{\end{equation}}
\newcommand{\ba}{\begin{eqnarray}}
\newcommand{\ea}{\end{eqnarray}}

\newcommand{\barr}{\begin{array}}
\newcommand{\earr}{\end{array}}

\newcommand{\refeqn}[1]{Eq. \eqref{#1}}
\newcommand{\refeqns}[1]{Eqs. \eqref{#1}}

\title[GR Simulations of Transonic Accretion Flows]{General Relativistic Numerical Simulation of sub-Keplerian Transonic Accretion Flows onto Black Holes: Schwarzschild Spacetime}

\author[J. Kim et al.]{
Jinho Kim,$^{1}$
Sudip K. Garain,$^{1}$\thanks{E-mail: sgarain@nd.edu}
Dinshaw S. Balsara$^{1}$
and Sandip K. Chakrabarti$^{2,3}$
\\
$^{1}$Department of Physics, University of Notre Dame, Notre Dame, IN 46556, USA\\
$^{2}$S. N. Bose National Center for Basic Sciences, Block JD, Sector III, Salt Lake, Kolkata, 700098, India\\
$^{3}$Indian Center for Space Physics, 43 Chalantika, Garia St. Rd., Kolkata, 700084, India
}

\date{Accepted XXX. Received YYY; in original form ZZZ}

\pubyear{2017}

\begin{document}
\label{firstpage}
\pagerange{\pageref{firstpage}--\pageref{lastpage}}
\maketitle

\begin{abstract}
We study time evolution of sub-Keplerian transonic accretion flows onto black holes using a general relativistic numerical simulation code. We perform simulations in Schwarzschild spacetime.
We first compare one-dimensional simulation results with theoretical results and
validate the performance of our code. Next, we present results of axisymmetric, two-dimensional simulation of advective flows. We find that
even in this case, for which no complete theoretical analysis is present in the literature, steady state shock formation is possible. 

\end{abstract}



\defcitealias{chakraba1989}{C89}
\defcitealias{chakraba1990}{C90}
\defcitealias{cm93}{CM93}
\defcitealias{mlc94}{MLC94}
\defcitealias{mrc96}{MRC96}
\defcitealias{ct95}{CT95}
\defcitealias{chakraba97}{C97}
\defcitealias{ggc2012}{GGC12}
\defcitealias{ggc2014}{GGC14}
\defcitealias{gc2013}{GC13}
\defcitealias{ggc2015}{GGC15}
\defcitealias{chakraba1996a}{C96a} 
\defcitealias{chakraba1996b}{C96b} 
\defcitealias{gcsr2010}{GCSR10}

\section{Introduction}

Independent of the source of matter supply, an accretion flow around a black hole is necessarily
transonic, i.e., it must pass through one or more sonic point(s). 
An accretion flow on a gravitating star,
whose specific angular momentum is everywhere or almost everywhere lower than that of the 
local Keplerian value may be termed as a sub-Keplerian flow. 
Theoretical calculations as well as numerical simulations of sub-Keplerian transonic accretion flows
around black holes have shown that such flows can have standing shocks for appropriate choice of the flow
parameters such as the specific energy $\epsilon$ and specific angular momentum $l$ of the accreting material
(\citealt{chakraba1989,chakraba1990}, hereafter C89, C90, respectively).
Paradoxical as it may sound, the flow decides to slow down just a few to few tens of gravitational
radii away from the horizon, simply because the centrifugal barrier becomes very strong
as compared to the gravity. However, the barrier is not unsurmountable and flow simply passes through a
shock transition to satisfy boundary conditions, as is normal in many astrophysical circumstances.
The shock can be simply standing or propagating away depending on the flow parameters and the
dissipative and cooling processes present in the flow 
(\citetalias{chakraba1989,chakraba1990}; \citealt{chakraba1996a}, and references therein).
Using numerical simulations with Smoothed Particle Hydrodynamics (SPH), these shocks have been shown to be 
stable in both one and two dimensional simulations
\citep[hereafter CM93, MLC94, respectively]{cm93,mlc94}.
They were found to propagate away for high enough viscosity \citep[][]{cm95}. 
The post-shock region, which is known as CENtrifugal pressure supported BOundary Layer (or CENBOL),
as in any other astrophysical flows, is found to extremely useful to explain
detailed spectral properties of the black hole accretion
disc quite satisfactorily \citep[hereafter CT95, C97, respectively]{ct95,chakraba97}.
The so-called Two-Component Advective Flow (TCAF) model, proposed in \citetalias{ct95} and \citetalias{chakraba97},
describes the most general structure of an accretion disc which consists of two transonic components,
namely, an optically thick (optical depth, $\tau\gg1$), geometrically thin Keplerian disc
and an optically thin ($\tau<1$), geometrically
thick sub-Keplerian flow. Shock is formed in the sub-Keplerian component which has
higher radial infalling velocity than the other component. In the post-shock region,
i.e., inside the CENBOL, these two components mix up due to turbulence and heat (\citetalias{ct95})
and form an optically slim ($\tau \sim 1$), geometrically thick
disc and continues its journey towards the central object. The CENBOL itself becomes responsible 
for the non-thermal power-law component of the observed spectrum.

Recently, TCAF model has been included in HEASARC's spectral analysis
software package XSPEC \citep{arn1996} and is being
used to study the spectral as well as timing properties of several black hole candidates
\citep{dcm2014,mdc2014,dmc2015,cmd2015,dmcm2015,jdcetal2016,cdcmj2016}.
It has been demonstrated that this model explains the observed data very well.
In outbursting sources, the shocks were found to be propagating as well - steadily moving towards the
black hole in the rising phase and moving away from it in the declining phase. 
Simulations of inviscid accretion flow with presence of 
a power-law cooling show that shock can oscillate about a certain mean
location, particularly when there is resonance between the cooling and the infall time scales.
{This may be considered to be an explanation of the origin of low
frequency quasi-periodic oscillations \citep[LFQPOs;][]{msc1996,cam2004}.}
Such oscillations have been shown to be present and stable when real Compton cooling
is present in the flow as well \citep[hereafter, GGC12, GGC14, respectively]{ggc2012,ggc2014}.
These intriguing properties of the CENBOL clearly demand conducting robust 
numerical experiments, which we set out to do in a series of papers.

Several numerical experiments are already present in the literature which study shock structures in
the sub-Keplerian flow around black holes. For one and two-dimensional inviscid, adiabatic flows,
it has been shown that a sub-Keplerian axisymmetric flow with and without shock is stable
(\citetalias{cm93}; \citetalias{mlc94}; \citealt{mrc96}, hereafter MRC96).
New codes have been tested against such non-linear solutions as well \citep[][]{toth1998}.
Numerical simulations of adiabatic viscous flow also demonstrate the
stability of these standing shock waves \citep{gc2012,lee2016}.
More recently, two-dimensional axisymmetric numerical simulations of viscous flow
in presence of power-law  and Compton cooling show that an advective flow actually splits into 
two components when appropriate viscosity parameters and cooling processes are chosen 
\citep[hereafter, GC13, GGC15, respectively]{gc2013,ggc2015}.
As the sub-Keplerian flow advects towards the central
black hole, angular momentum transport and condensation due to cooling on the equatorial plane help
the flow to segregate into two distinct advective components, each being separately transonic.
The component near the equatorial plane has been shown to have the
properties very similar to a standard Keplerian disc \citep{ss73}.
This component is surrounded by the sub-Keplerian component which will have the steady shock. Thus,
these results show that a stable TCAF formation is indeed possible.

{In case of magnetized accretion disc, magnetorotational instability 
\citep[MRI;][]{balbus1991, hawley1992} can make the disc turbulent and
this turbulence may transport angular momentum outwards efficiently. 
However, it has been shown that turbulence triggered by MRI produce the value
of the alpha viscosity parameter $\sim$ 0.01
\citep{brand1995, hawley1995, hawley1996, arlt2001, smak1999, king2007, kotko2012}.
In the literature, many authors from various groups published results of analytical
viscous solutions as well as viscous hydro-dynamic simulations
where it has been shown the shock is stable if the viscosity parameter is lower than
a critical viscosity parameter 
\citep{chakraba1990b, chakraba1996d, cm95, lmc1998, lcscbz2008, lee2011, das2014, lee2016}.
Also, there have been many stability studies of shock
\citep{nakayama1992, nakayama1994, nobuta1994, gu2003, gulu2006}, 
but it was shown that even under non-axisymmetric perturbations, the shock
tends to persist, albeit, as a deformed shock \citep{mtk1999}.
}

Despite all these important developments, almost all
the above mentioned simulations have been performed using the so-called
pseudo-Newtonian potential \citep{pw1980} which mimics the Schwarzschild spacetime.
This potential retains the particle properties of the Schwarzschild geometry in the sense that the 
marginally bound and marginally stable orbits are located at exactly the same places as in GR. However, several 
properties in the strong gravity limit just outside of the horizon are not retained with precision.
For instance, the energy released at the marginally stable orbit, or the velocity of matter on the horizon
are different. Thus, though the results were satisfactory, it was difficult to judge if
any of the effects observed were artefacts of the potential. Fortunately, even in general relativistic (GR) framework, in Kerr spacetime, such shocks have been shown to exist
(\citetalias{chakraba1990}; \citealt[][hereafter C96b]{chakraba1996b}).
And it became clearer that the shocks in black hole accretion are indeed possible only because of the presence of the 
inner sonic point between the marginally bound and marginally stable orbits where strong gravity is important. 
{Strong gravity forces the flow to have an inner sonic point. Most importantly
in accretion flow configurations, the solution passing through the inner sonic point
has higher entropy than that passing through the outer sonic point. So the flow
generates entropy at the shock and then passes through the inner sonic point. The
location of the sonic point is the closest indicator of a stable fluid (unlike the marginally
stable orbit, which is the closest indicator for a particle trajectory).
}
Thus there are all the more reasons to verify these results using a full relativistic framework.
In this work, we perform a general relativistic simulation of the sub-Keplerian flow in 
Schwarzschild spacetime. To our knowledge, no numerical experiment has so far been performed 
which tests the possibility of a stable CENBOL formation in general relativity.

This paper is organized as follows. In Section \ref{sec:sec1}, we present the analytical method
to calculate the one-dimensional flow properties in Schwarzschild spacetime. In Section
\ref{sec:sec2}, we present the general relativistic equations which are solved numerically and
the numerical procedure we use for doing this. In the next Section, we present the results for one and two dimensional
simulations. Finally,  we present our conclusions.

In this paper, we choose $R_g=GM_{\rm{BH}}/c^2$ as the unit of distance,
$R_g c$ as unit of angular momentum, and $R_g/c$ as unit of time.
In addition, we choose the geometric units $G=M_{\rm{BH}}=c=1$
($G$ is gravitational constant, $M_{\rm{BH}}$ is the mass of the black hole and
$c$ is the unit of light). Thus $R_g=1$, and angular momentum and time are
measured in dimensionless units.

\section{Analytical Solution}
\label{sec:sec1}

The general relativistic study of transonic flows has been done extensively
by Chakrabarti (\citetalias{chakraba1990,chakraba1996b}). Therefore, we do not describe
all the details here. However, for completeness, we only mention the important 
equations for Schwarzschild spacetime.

For this calculation, we use Boyer-Lindquist coordinates ($t,r,\theta,\phi$).
The line element in Schwarzschild spacetime is given as follows,
\begin{equation}
\label{eqgmunu}
\begin{split}
ds^2&= g_{\mu\nu}dx^{\mu}dx^{\nu}\\
&= - \left(1-\frac{2}{r}\right)dt^2 + \left( 1-\frac{2}{r}\right)^{-1}dr^2
+ r^2 d\theta^2 + r^2 \sin^2\theta d\phi^2
\end{split}
\end{equation}
We are interested in the flow close to the equatorial plane, so $\theta=\pi/2$ is assumed for
analytical study.

In absence of viscosity and any heating or cooling, one can find the 
conserved specific energy as (\citetalias{chakraba1996b}) 
\begin{equation}
\label{eqE}
\epsilon =hu_t=\frac{1}{1-na^2}u_t,
\end{equation}
where, $n=1/\left(\Gamma-1\right)$ is the polytropic index, $\Gamma$ being the adiabatic index
and  $h=1/\left(1-na^2\right)$ is the enthalpy, $a$ being the sound speed. Also,
\begin{equation}
u_t=\left[\frac{1-\frac{2}{r}}{(1-V^2)(1-\Omega l)}\right]^{1/2}.
\end{equation}
Here,
\begin{equation}
\label{omega}
\Omega = \frac{u^\phi}{u^t}=-\frac{lg_{tt}}{g_{\phi\phi}} = \frac{l}{r^2}\left(1-\frac{2}{r}\right),
\end{equation} 
and $l=-u_\phi/u_t$ is the specific angular momentum. 
Also, 
\begin{equation}
\label{bigV}
V=\frac{v}{\left(1-\Omega l\right)^{1/2}},
\end{equation} 
where 
\begin{equation}
\label{smallv}
v=\left(-\frac{u_ru^r}{u_tu^t}\right)^{1/2}.
\end{equation}

The entropy accretion rate (\citetalias{chakraba1989,chakraba1996b}) is given by
\begin{equation}
\label{eqM}
\dot{\mu}=\left(\frac{a^2}{1-na^2}\right)^n V \left(1-\Omega l\right)^{1/2} u_t r^2
\end{equation}

We follow the usual solution procedures use in transonic flows (\citetalias{chakraba1989,chakraba1990}) to calculate
$V(r)$ and radial dependence of other required quantities. By differentiating equations (\ref{eqE}) 
and (\ref{eqM}) with respect to $r$ and eliminating terms involving $da/dr$, we find following
expression as the gradient of $V(r)$:
\begin{equation}
\label{eqdVdr}
\frac{dV}{dr}=\frac{V\left(1-V^2\right)\left[1-2ra^2+3a^2-\frac{\Omega l}{1-\Omega l}\left(r-3\right)\right]}{r\left(r-2\right)\left(a^2-V^2\right)}
\end{equation}

We can readily see that for $l=0$, we recover the similar expression for Bondi flow
onto a black hole (Eq. 1.29 of \citetalias{chakraba1990}). At the sonic point, both numerator
and the denominator vanish and one obtains the so-called sonic point condition as
\begin{equation}
\label{soniceqn}
\begin{split}
V_c&= a_c, \\ {\rm and} \\
a_c^2&= \frac{1}{2r_c-3}\left[ 1-\frac{\Omega l}{1-\Omega l}\left(r-3\right) \right],
\end{split}
\end{equation}
where, $r_c$ is called the sonic radius.

To find a complete solution from the horizon to infinity, one needs to supply the
specific energy $\epsilon$ and the specific angular momentum $l$. If the supplied
parameters allow the accretion solution to have a shock in the sense that Rankine Huguniot
conditions are satisfied, then the shock location can
be found by determining a constant, $C$, which remains invariant across the shock
(\citetalias{chakraba1989,chakraba1990}). It is found that this invariant quantity is the 
same in Schwarzschild space-time and in pseudo-Newtonian potential 
\citepalias{chakraba1990}, as this is a local equation. Therefore, we use the following expression,
\begin{equation}
C=\frac{\left[\Gamma M + \left(1/M\right)\right]^2}{2+\left(\Gamma-1\right)M^2}
\end{equation}
to determine the shock location for this calculation \citep{cd2001}.
Here, we use $M=V/a$ as the definition of Mach number.

\section{Numerical Simulation Procedure}\label{numer}
\label{sec:sec2}

We use so called ``Valencia formulation'' to numerically solve the relativistic hydrodynamic equation \citep{ban1997}.
This formulation gives flux conservative form of the system of hydrodynamics equation in the framework of 3+1 formalism.
It has been applied very successfully in computational fluid dynamics.
In our coordinate system, the conservative variables ($q$) and primitive variables ($w$) are
\be
q=\left(
\begin{array}{c}
D \\ S_r \\ S_{\theta} \\ S_{\phi} \\ \tau
\end{array}
\right)
{\equiv}\left(
\begin{array}{c}
\rho W \\ \rho h W^2 v_r \\ \rho h W^2 v_{\theta} \\ \rho h W^2 v_{\phi} \\ \rho h W^2 -P -D
\end{array}
\right),\,
w=\left(
\begin{array}{c}
\rho \\ v^r \\ v^{\theta} \\ v^{\phi} \\ P
\end{array}
\right).
\label{eq11}
\ee
Here $\rho$ is the fluid rest mass density $P$ is the pressure and $h$ is the specific enthalpy. 
They are measured in the comoving frame of the fluid.
$v^i$ is the fluid velocity measured by Eulerian observer.
$W$ is the Lorentz factor and defined as $W=1/\sqrt{1-\gamma_{ij} v^i v^j}$.
Here, $\gamma_{ij}$ are the spatial part of the metric components $g_{\mu\nu}$.
The radial and angular velocity in the Eulerian frame can be expressed in 
terms of $v$ in \refeqn{smallv} and $\Omega$ in \refeqn{omega}:
\be\label{vrvphi}
\begin{split}
v^r&={\frac{u^r}{W}}=\left(1-\frac{2}{r}\right)^{\frac{1}{2}} v \\
v^{\phi}&={\frac{u^{\phi}}{W}}=\left(1-\frac{2}{r}\right)^{-\frac{1}{2}} \Omega
\end{split}
\ee

Assuming axisymmetry ($\frac{\partial}{\partial\phi}=0$), the hydrodynamical equations in the curved spacetime that described in \refeqn{eqgmunu} can be written as follows:
\be
\frac{\partial\left(\sqrt{\gamma}q\right)}{\partial t}
+\frac{\partial\left(\sqrt{-g}f^r\right)}{\partial r}
+\frac{\partial\left(\sqrt{-g}f^{\theta}\right)}{\partial \theta}
=\sqrt{-g}\Sigma,
\label{eq13}
\ee
where
\ba
f^r&=&\left[
\begin{array}{ccccc}
Dv^r\\S_rv^r+P\\S_{\theta}v^r\\S_{\phi}v^r\\\tau v^r+P v^r
\end{array}
\right],\nonumber\\
f^{\theta}&=&\left[
\begin{array}{ccccc}
Dv^{\theta}\\S_rv^{\theta}\\S_{\theta}v^{\theta}+P\\S_{\phi}v^{\theta}\\\tau v^{\theta}+Pv^{\theta}
\end{array}
\right],\nonumber\\
\Sigma&=&\left[
\begin{array}{ccccc}
0\\
-\frac{\rho h W^2}{r}\left(\frac{1}{r-2}\left(1+v_r v^r\right) - v_\theta v^\theta - v_\phi v^\phi \right)+\frac{2P}{r} \label{eq14-1}\\
\cot\theta\left(\rho h W^2 v_\phi v^\phi + P\right)\label{eq14-2}\\
0\\
-\frac{\rho h W^2 v^r}{r\left(r-2\right)}\label{eq14-3}
\end{array}
\right]. \label{eq14}
\ea
Here $\sqrt{\gamma}$ and $\sqrt{-g}$ are the determinants of spatial and spacetime metric, respectively.
From the Schwarzschild metric shown in \refeqn{eqgmunu}, we have $\sqrt{\gamma}=r^2\sin\theta\left(1-2/r\right)^{-1/2}$, and
$\sqrt{-g}=r^2\sin\theta$.
For the spherically symmetric cases, $\frac{\partial}{\partial\theta}=0$.
{\refeqn{eq14} consists of the continuity, three momentum and energy equations.
We clearly see the conservation of total rest mass (baryon number) and angular momentum
in the first and fourth rows of \refeqn{eq14}. Particularly, the terms of $\Sigma$ in the
momentum equation contains the gravitational and the centrifugal forces.
The gravitational force which is purely radial in the Schwarzschild metric
is shown in the first term of the second row of $\Sigma$. The second and
third terms in the second row as well as the first term in the third row
represents the centrifugal force by the rotation velocity. Since the
centrifugal force, exerted by the $v^\phi$, is not purely radial, it contributes
to both $r-$ and $\theta-$ momentum equations. (c.f. The centrifugal force
exerted by the $v^\theta$ is purely radial. Therefore, it only appears in the
radial momentum equation.) Note that the last terms in the $\Sigma$ of
the momentum equations are the additional terms in the spherical polar
coordinates system. The fifth row of $\Sigma$ is the sink or source of the
energy contributed by gravity.}

We use the ideal gas equation of state which can be written in the following form:
\be\label{eq30}
P=\left(\Gamma-1\right)\rho e,
\ee
where $e$ is the specific internal energy.
The above equation of state provides the expression of specific enthalpy:
\be\label{findP}
h=1+\frac{\Gamma}{\Gamma-1}\frac{P}{\rho}.
\ee 

In this paper, we solve the above hydrodynamic equations using a numerical code developed by 
\citet{kim2012}.
The details of the code can be found in \citet{kim2012}.
The most useful property of this code is that it can be applied to any spacetime 
metric in any coordinate system. This code uses finite volume methods to ensure 
local conservation of the fluid in the computational grid.
Therefore, the code can guarantee total mass and angular momentum conservations 
which appear in the first and fourth rows of \refeqn{eq14}.
For the treatment of the discontinuous behaviors of the fluid such as shocks, 
rarefactions or contact discontinuities, the High Resolution Shock Capturing (HRSC) 
techniques are applied in the code.
We use the third order slope limiter proposed by \citet{shi2003} which is based 
on the {\it minmod} function. For the flux approximation, we use the HLL method \citep{har1983}. 
The HLL method has some dissipation but the results are very stable.
For the time integration, we use the third order three stage Strong Stability-Preserving (SSP) 
Runge-Kutta method which is known as Shu-Osher method \citep{shu1988}.

\section {Results}
\label{sec:sec3}

We use inflow boundary condition at the outer boundary located at $r_{\rm{out}}=100$. 
The inner boundary is placed at $r_{\rm{in}}=2.1$ and we use the extrapolated values of the primitive variables for the inner ghost cells.
These inner ghost cells are located outside of the event horizon. 
In this paper, we present results of one-dimensional 
and two-dimensional simulations. For better resolution close to the central
black hole, we logarithmically binned the radial direction in 300 zones for
all the simulation results presented here. For 
two-dimensional simulations, in addition to above mentioned radial binning, 
we used 100 equi-spaced zones in polar direction.
The innermost radial zone ($\Delta r$) has a size of $2.72\times10^{-2}$ and the outermost zone has a size of 1.28.

For these simulations, we need values of the primitive variables (see, \refeqn{eq11})
at $r=r_{\rm{out}}$. For all the simulations, we set $v^\theta=0$. 
The density of the incoming matter at $r_{\rm{out}}$ is normalized to $1$.
In absence of self-gravity and heating or cooling, the density is scaled out 
and the simulation results remain valid for any accretion rate
(\citetalias{mrc96}; \citealt{gcsr2010}, hereafter GCSR10).
We evaluate $V(r=r_{\rm{out}})$ by solving \refeqn{eqdVdr} for a given pair of conserved flow variables.
Subsequently, the sound speed $a_{\rm{out}}=a(r=r_{\rm{out}})$ is evaluated using \refeqn{eqE}.
Next, using \refeqns{bigV}, (\ref{smallv}) and (\ref{vrvphi}), we evaluate $v^r(r=r_{\rm{out}})$ and
$v^\phi(r=r_{\rm{out}})$. Pressure of the incoming matter is evaluated from the sound speed 
$a_{\rm{out}}$ using \refeqn{findP}. These values are maintained in the ghost zones
of the outer boundary of our computational domain.
As an initial condition, we put floor values for the density  to be $\rho_{\rm{floor}}=10^{-10}$ (in normalized unit)
and the corresponding floor value of pressure inside the computational domain. 
{Floor value of the pressure is chosen such that the sound speed (or temperature) of
the background matter is same as that of incoming matter (\citetalias{mrc96,gc2013}).
Thus, once we know $a_{\rm{out}}$, we evaluate the value of the pressure floor using 
\refeqn{findP} by substituting $\rho=10^{-10}$ in this equation.
Then, the value of pressure floor is $P_{\rm{floor}}=na_{\rm{out}}^2\rho_{\rm{floor}}/\left[\left(1-na_{\rm{out}}^2\right)\left(n+1\right)\right]$.}
Initially, the velocity components in the
floor grids are set to zero. Thus, initially, as the matter rushes towards the black hole,
it fills the vacuum rapidly. After the matter reaches the inner boundary, the flow starts to feel the
pressure and centrifugal force.

\subsection{One-dimensional, spherically symmetric Bondi accretion}

For a given $\epsilon$, the analytical structure of a spherically symmetric
Bondi accretion flow is completely determined ($l=0$ for this flow). 
For the comparative study, we choose $\epsilon=1.015$. This choice gives $V=0.053$ 
and $a=0.088$ at $r_{\rm{out}}$, and the sonic point at $r=54.04$. Note that, just the energy was
sufficient to determine all the quantities as the other conditions come from transonicity (\refeqn{soniceqn}; \citetalias{chakraba1990}).

\begin{figure}
\includegraphics[width=\columnwidth]{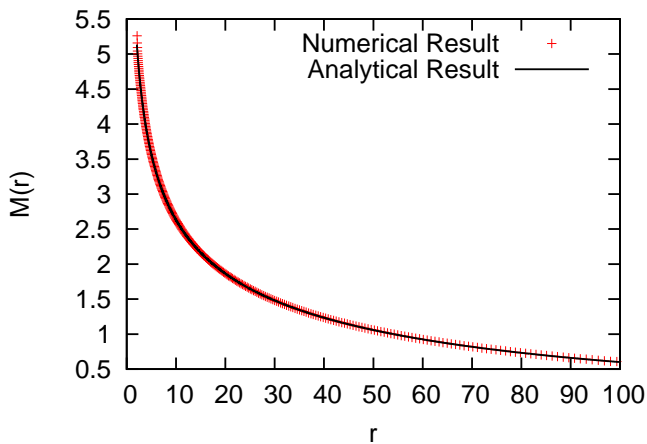}
\caption{ Comparison of radial Mach number variation for Bondi
accretion flow. Black solid line is obtained using analytical method,
whereas red plus signs are the simulation results after steady state is reached. They appear to be
indistinguishable.}
\label{fig1}
\end{figure}
In Fig. \ref{fig1}, we compare the radial variations of Mach number, defined as $M(r)=V(r)/a(r)$. 
The black solid line represents the analytical result, whereas the red plus signs represent 
the numerical simulation result. Clearly, they are indistinguishable. Dynamical time, $t_{\rm{dyn}}$
(defined as the time required for matter to reach inner boundary from $r_{\rm{out}}$ 
in steady state) is found to be $t_{\rm{dyn}}\sim991$ for this simulation and we 
ran the simulation for more than $20 t_{\rm{dyn}}$ ($t\sim19800$).
Also, we ran some more cases by varying the number of radial zones ranging from 150 to 600 and verified that the results remained converged.
Note that the inner boundary places at $r_{\rm{in}}=2.5$ for the lower resolution case in order to prevent the inner ghost cells from locating inside the event horizon. 

\subsection{One-dimensional accretion flow with non-zero angular momentum}\label{sec:1d_result}

When angular momentum is present, the flow structure changes significantly
depending on its strength. As discussed earlier, depending on the values of 
$\epsilon$ and $l$, the accretion flow may or may not have any shock. In this
Section, we present the results of two simulations in one-dimension.
One solution has a shock and the other does not. 
As discussed in \citetalias{chakraba1989} {(see Fig. 3)};
 \citetalias{chakraba1990} {(see Chap. 3 and 6)}, the accretion
solution having a shock first passes through the outer sonic point, makes
a transition to the sub-sonic branch at the shock and then passes through
the inner sonic point just before being supersonically accreted by the black hole. On the other hand,
when $\epsilon$ and $l$ are such that it is not possible to have a shock, the flow
passes only through the outer or inner sonic point before disappearing behind the horizon.
However, such a flow still can slow down as it approaches the black hole because of the
centrifugal force.

\begin{figure}
\includegraphics[width=\columnwidth]{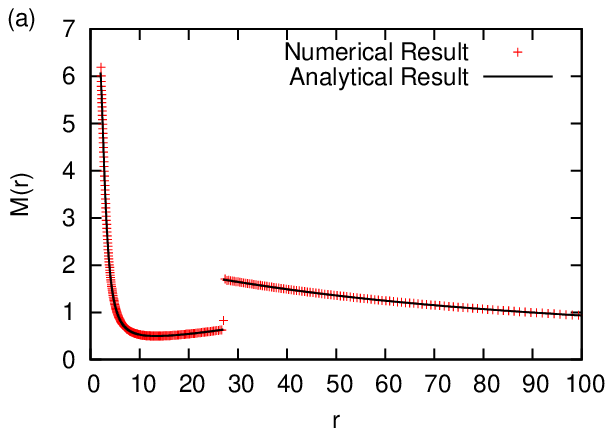}
\includegraphics[width=\columnwidth]{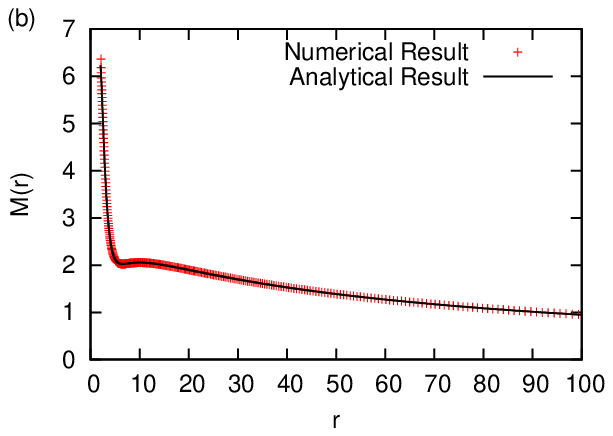}
\caption{Comparison of radial Mach number variation for one-dimensional axisymmetric
accretion flows. (a) represents results where shock is present and (b) 
represents results where no shock formed and the flow is supposed to pass only through the 
outer sonic point. Black solid line is obtained using analytical method,
whereas red plus signs are the simulation results.}
\label{fig2}
\end{figure}
In Fig. \ref{fig2}(a), we present the radial variation of Mach number, $M (r)$, 
at the final steady state for the accretion
flow which has a shock. As before, the black solid line represents the analytical result 
and the red plus signs
represent the simulation result. $\epsilon=1.007$ and $l=3.4$ are chosen for this simulation.
Analytical calculation gives $V=0.0645$ and $a=0.0691$ at $r=100$.
The outer sonic point, shock and inner sonic point are located at $r=89.59$, $26.98$
and $5.38$ respectively. As can be seen, our 1D simulation has captured
all the locations properly. For this case, dynamical time is found to be 1150 and
the simulation was run till $20 t_{\rm{dyn}}$ ($t\sim23000$). We have also computed $\epsilon$ and $l$ from the simulation
result and verified that these two quantities remain constant along the flow.

Fig. \ref{fig2}(b) shows the radial variation of $M$ for a case where there is no shock in the
accretion solution, and is expected to pass only through the outer sonic point. 
We choose $\epsilon=1.007$ and $l=3.25$ to compute the outer
boundary values for this simulation. For this case, dynamical time is found to be 920 and
the simulation was run till $20 t_{\rm{dyn}}$ ($t\sim18400$). The outer sonic point is located at $r=91.92$
for these parameters. As can be seen from the Figure, analytical and numerical results
are nearly inseparable and hence, we see that numerical simulation has captured
this location very well. We may mention in passing that the simulations presented in the literature are
primarily for supersonic injection to save computational time. However, our efficient method allows us to inject 
the flow at sub-sonic Mach number.

\begin{figure}
\includegraphics[width=\columnwidth]{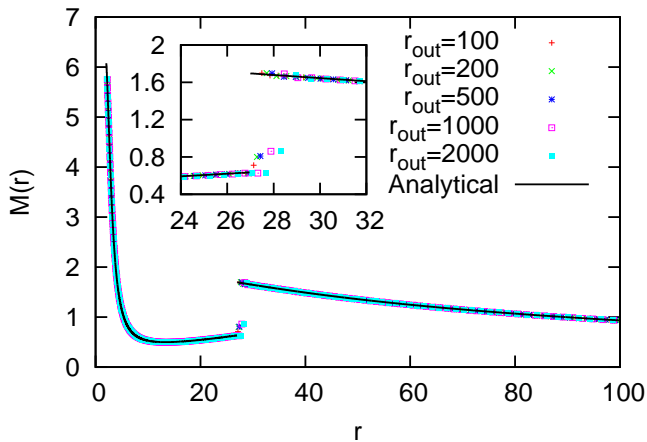}
\caption{ Results of rerun for the case presented in Fig. \ref{fig2}(a) 
by moving $r_{{\rm out}}$ to 200, 500, 1000 and 2000. Green, blue, magenta and cyan (cross, star, open squrare and filled square) 
points represent the simulation results for $r_{{\rm out}}=200,~500,~1000{\rm ~and}~2000$ 
respectively. We can clearly see that the shock locations do not get affected
if we change $r_{\rm{out}}$. In the inset, we show the zoomed in part around 
the shock location.}
\label{fig2_add}
\end{figure}

{In order to verify that the shock location is not affected by the outer boundary,
we rerun the case presented in Fig. \ref{fig2}(a) by moving $r_{{\rm out}}$ to 
200, 500, 1000 and 2000.
All these simulations have been run using 300 zones in the radial direction.
In Fig. \ref{fig2_add}, we present these simulation results. Green, blue, magenta
and cyan (cross, star, open squrare and filled square) points represent the simulation results for $r_{{\rm out}}=200,~500,~1000{\rm ~and}~2000$ respectively.
In the inset, we show the zoomed in part around the shock location.
Comparison of shock location for various $r_{\rm{out}}$ shows that it
does not get affected if we change $r_{{\rm out}}$. 
Slight mismatch for $r_{\rm{out}}=1000$ and $2000$ may be due to the grid
size variation at the shock location. }

\subsection{Two-dimensional simulations}

Realistically, an accretion disc is three dimensional. However, 
assuming axisymmetry we can study the disc structure in two-dimensions. On the other hand, the 
theoretical formalism of transonic flows \citep[][]{chakraba1996b}, is developed for flows in vertical equilibrium
which is quite thin. Indeed, the shock location and Mach number variation depend on the model assumption. Most 
interestingly, it was shown (\citetalias{gcsr2010})
that the pre-shock flow behaves as a conical flow, while the post-shock behaves as a flow in vertical equilibrium.
So a direct comparison with theoretical result is not possible in a two dimensional 
simulation. On the other hand, in \autoref{sec:1d_result}, we showed that our simulation result matches
with the theory very well. Thus the results obtained in two dimensions may be trusted. 

For the results presented here, we do the simulation in $\left(r, ~\theta\right)$ coordinates. 
Simulation domain extends from $r_{\rm{in}}=2.1$ to $r_{\rm{out}}=100$ in radial direction 
and $[0:\pi/2]$ in polar direction. The incoming matter enters the simulation box
at $r_{\rm{out}}$ through one-tenth of polar zones starting from equatorial plane
(\citetalias{gcsr2010,ggc2012,ggc2014}). For this simulation, we evaluate the outer
boundary values using vertical equilibrium model. We choose $\epsilon=1.0022$ and $l=3.36$,
and this choice gives $V=0.0825$ and $a=0.0526$ at $r_{\rm{out}}$. For these parameters,
the one-dimensional analytical calculation predicts the shock to be located at $15.77$. However,
in a full 2D simulation, presence of turbulence due to centrifugal barrier 
is expected to shift the shock farther out. 
{As the incoming matter hits the centrifugal barrier, some matter bounce back and
interact with the incoming matter. Thus a turbulence is generated. This turbulent
pressure seems to be comparable to the other pressure effects, such as thermal
and ram pressure \citepalias{mlc94,mrc96}.
This turbulent pressure shifts the location of the shock further out compared to the
location that we calculate using one-dimensional analytical method.}
Interestingly this is precisely what we see.

\begin{figure*}
\begin{center}
\begin{tabular}{cc}
\includegraphics[width=\columnwidth]{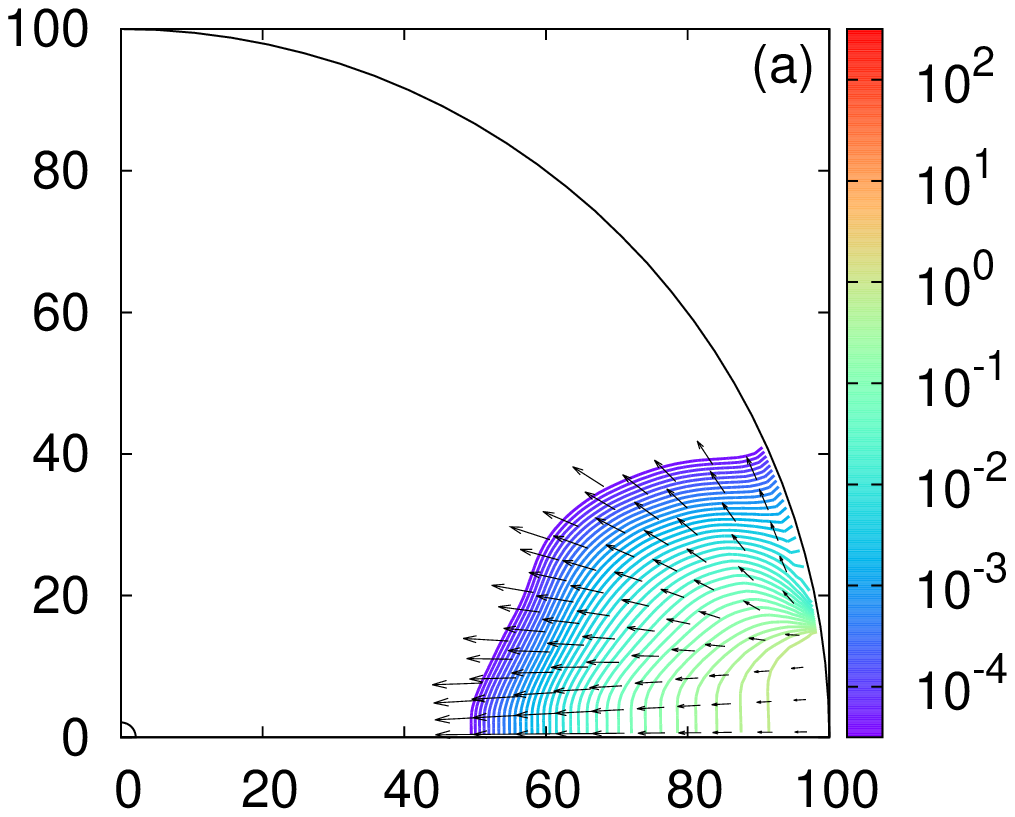} &
\includegraphics[width=\columnwidth]{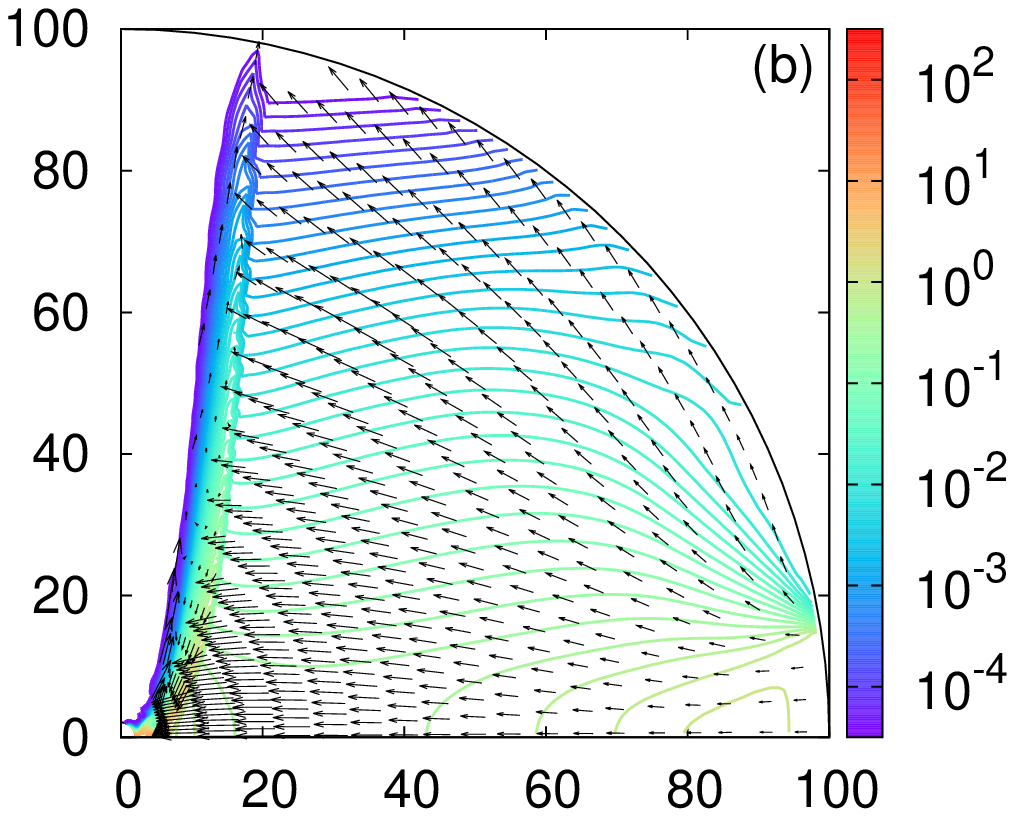} \\
\\
\includegraphics[width=\columnwidth]{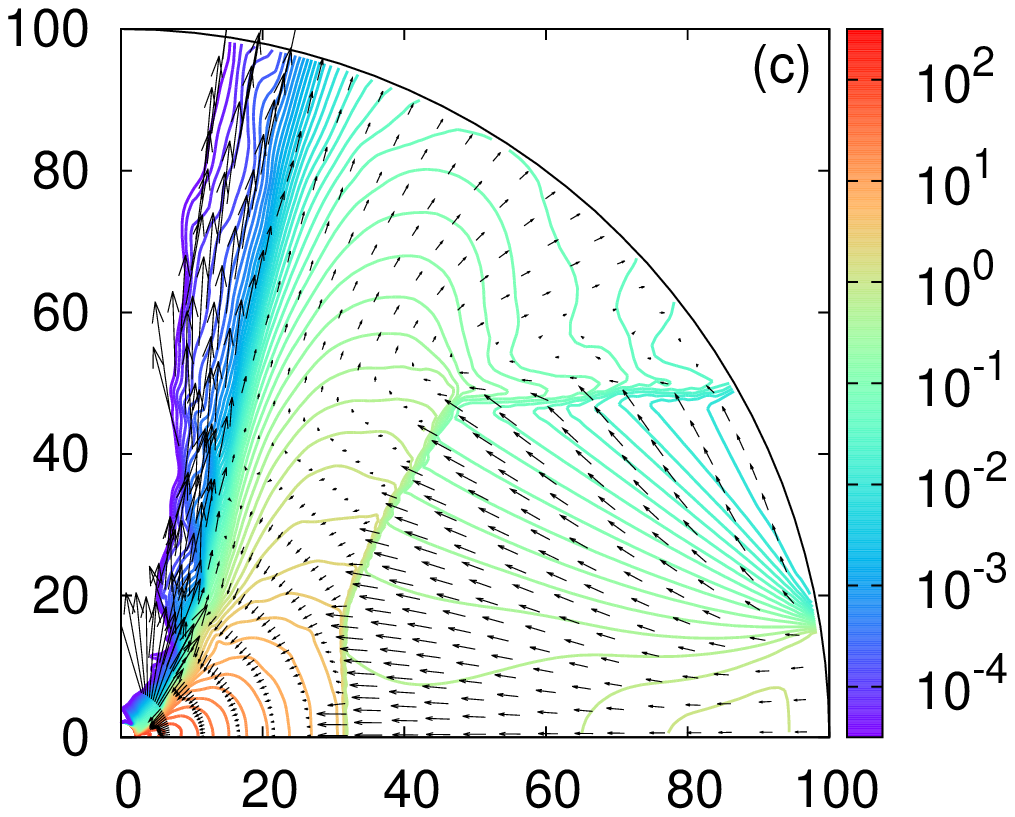} &
\includegraphics[width=\columnwidth]{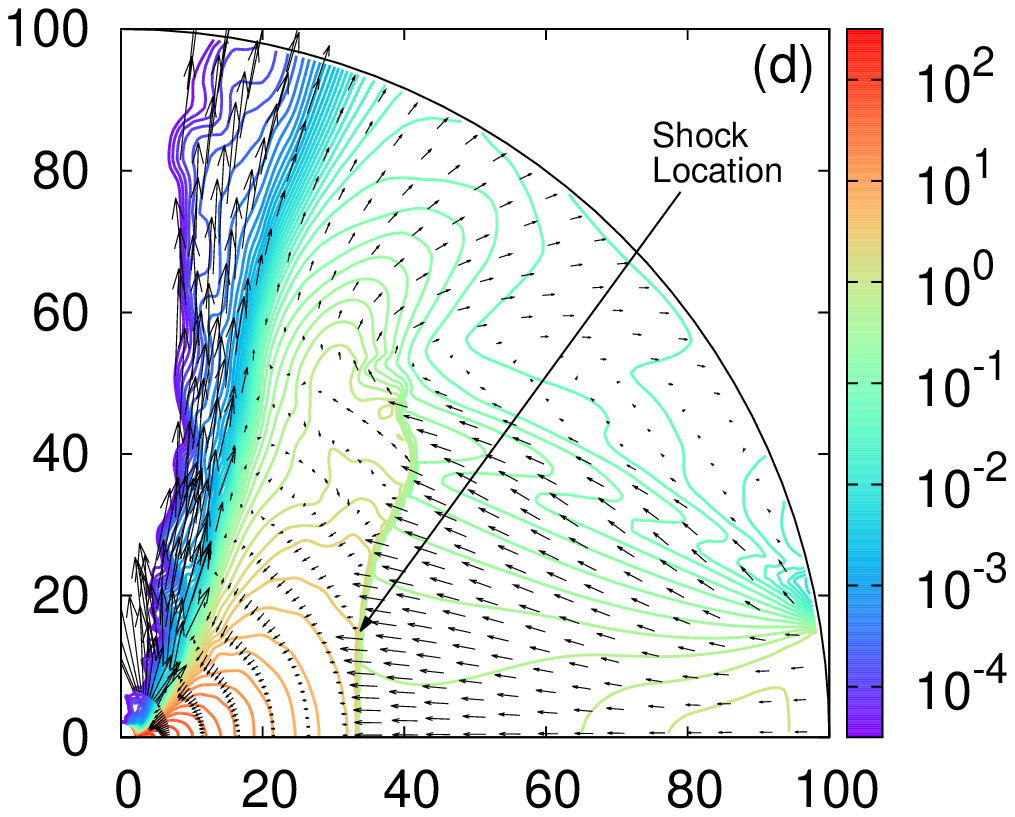}
\end{tabular}
\caption{Contours of normalized density, over-plotted with velocity
vectors at four different time. Snapshots (a), (b), (c) and (d) are plotted at times 158,
496, 2250 and 30000, respectively. The corresponding dynamical times are 0.11, 0.33, 1.5 and 20, respectively. See text for details.}
\label{fig3}
\end{center}
\end{figure*}

In Fig. \ref{fig3}, we present the contours of normalized density inside the 
accretion disc at four different times. Associated color bar represents
the values of the normalized density.
Velocity vectors are over-plotted with density. The length of a vector
is proportional to the magnitude of velocity at that location. 
The simulation is carried out till the time of 30000 and we measure the dynamical
time to be $t_{\rm{dyn}}\sim1500$. Thus, the simulation continued for at least 20 dynamical
times without any significant time evolution and hence, we believe that the system has reached a steady state.

\begin{figure}
\includegraphics[width=\columnwidth]{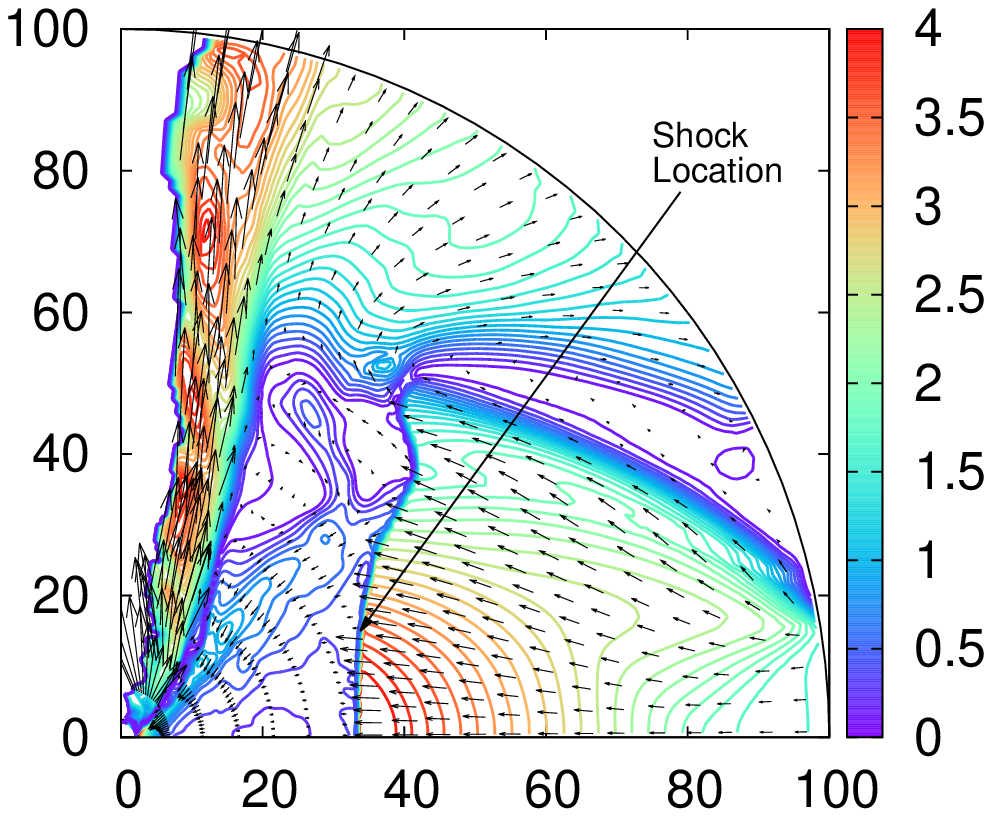}
\caption{Mach number contours at the final time, over-plotted with
velocity vectors. It can be seen that the high velocity outflow leaves
the computational domain super-sonically.}
\label{fig4}
\end{figure}

Fig. \ref{fig3}(a) shows the density and velocity vectors at time 158 ($t_{\rm{dyn}}\sim0.11$), a little after 
the simulation started. Fig. \ref{fig3}(b) shows the same at time 496 ($t_{\rm{dyn}}\sim0.33$), soon after matter 
reaches $r_{\rm{in}}$. It can be seen that a shock structure already started forming 
by this time. The shock front can be identified by the jump in density color contour
close to the axis and also a sudden reduction of velocity vector lengths. 
The contours clearly look like those of a thick accretion disc as discussed in \citetalias{mlc94} also for a 
pseudo-Newtonian simulation.
As we move up in the vertical direction from the equatorial plane, 
the shock front bends outward. This is explained in \citetalias{mrc96}; \citealt{chakraba1996c} as due to the reduction of 
gravitational pull with height, while the  centrifugal force which remains almost the same, pushes the shock
outward.
This shock front moves away radially from the black hole and stabilizes at a radial 
distance of $\sim33.8$. This can be seen in Fig. \ref{fig3}(c) which is a snapshot at 
nearly 1.5 dynamical times, i.e., 2250. Fig. \ref{fig3}(d) shows the snapshot at a final 
time of 30000 ($t_{\rm{dyn}}\sim20$). There appears to be practically no change in the shock 
location or the flow behaviour at this period of the simulation. 
We ran this simulation at $600\times200$ resolution and found that the results presented here are converged. We also ran this simulation at $150\times50$ resolution by placing $r_{\rm{in}}$ at 2.5. Even for this lower resolution, the shock location and the overall structure of the disc is found to be very similar to the presented results. 
Velocity vectors in Figs. \ref{fig3}(c) and \ref{fig3}(d)
also show that a strong outflow emerges from the post-shock region or CENBOL. 
In order to prove that it indeed leaves the system supersonically, we plot in 
Fig. \ref{fig4} the contours of radial Mach number. As the color code would indicate,
the inward flow became highly supersonic in the pre-shock region and highly subsonic
in the post-shock region. It became supersonic again closer to the black hole as it moves to satisfy the 
boundary condition on the horizon. The outflow behaves exactly the opposite way. 
It starts subsonically from CENBOL surface and becomes supersonic by the time it reaches
at about ten Schwarzschild radii. The raggedness of the contours on the funnel like 
vortex surface close to the axis is due to the artefact of finite sized grids whose size increase
with radial distance from the black hole.

\begin{figure}
\includegraphics[width=\columnwidth]{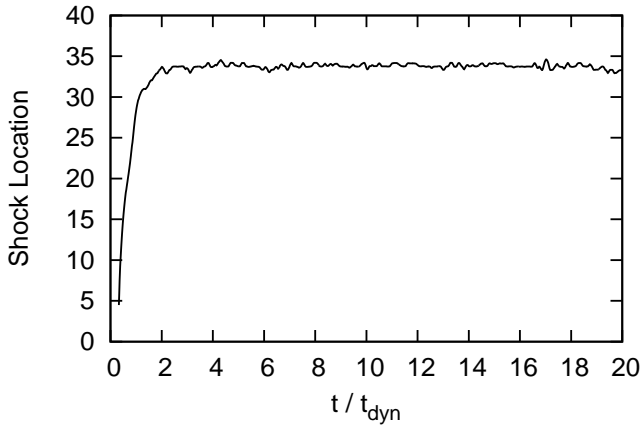}
\caption{Time variation of shock location. The shock structure is steady, 
although it oscillates slightly about its mean location of $\sim33.8$}.
\label{fig5}
\end{figure}
In Fig. \ref{fig5}, we present the time evolution of the shock location close to 
the equatorial plane. We dynamically compute the shock location by 
picking up the position where the Mach number changes from $M>1$ to $M<1$. 
We have omitted the initial transient time from this plot when the floor grids were being filled
by the incoming flow for the first time.
Fig. \ref{fig5} shows that the shock location has achieved a steady state, although
it oscillates slightly about its mean location of $r\sim33.8$, possibly due to the finite grid size at this 
distance.
\section{Conclusions}
\label{sec:sec4}

It has been demonstrated that TCAF model explains spectral and temporal
properties of the black hole accretion discs very well. The CENBOL region which
is primarily responsible for emission of hard photons and the outflows, relies on the formation
of shock in the sub-Keplerian component in this model. However, the formation of CENBOL
was taken for granted only using simulations carried out in pseudo-Newtonian geometry
although its genesis lies in properties of an advective flow in the strong gravity limit.
In this paper, we performed numerical simulations in fully general relativistic 
framework in Schwarzschild spacetime to see whether
a steady shock formation is still possible or not. We presented results of one-dimensional
and two-dimensional simulations. For one-dimensional simulations, we tested the code by comparing 
our results with the steady state results obtained using analytical methods. 
We ran our code for sufficiently long time 
(over 20 dynamical time) the steady state was found to be achieved.
Comparisons of radial Mach number variations were shown for spherically 
symmetric Bondi accretion flow as well as for axisymmetric
accretion flow with and without shock. We found a very good match between the 
two methods even at moderate resolution and these validated the performance 
of our simulation code. This gave us confidence to  delve into unchartered 
territory of running the code in two-dimensions, where, strictly speaking, a fully self-consistent
theoretical result was missing.  Here again, we found steady shock formation and the post-shock
region can be easily identified to be the CENBOL region used by \citetalias{ct95}. 
We found that a centrifugal force driven supersonic jet is formed from the surface of the 
CENBOL. We also plotted the time variation of the shock location close to the equatorial plane.
We found that the location slightly oscillates about a mean value of $\sim 33.8$, but overall  structure is steady.

The results presented here were obtained from purely hydrodynamical simulations
in Schwarzschild spacetime. Analytically, it has been shown in \citetalias{chakraba1990,chakraba1996b} that the
presence of black hole rotation will affect the structure of an accretion disc. 
The parameter space of the shock formation is different for prograde and retrograde
flows, and locations of the sonic points as well as shock can be closer or
away depending on whether the so-called ``spin-orbit'' coupling term (arising out of the 
product of the spin vector  of the black hole and the orbital angular momentum vector  of the matter) 
is positive or negative.
We will numerically study the effects of black hole rotation on the accretion disc and shock
behaviour. Our code would be easily used to study the dragging of inertial effects also which is 
expected to affect the outflows and CENBOL behaviour.
Furthermore, in presence of dissipations, such as viscosity and/or heating/cooling, 
the structure and the location of the CENBOL will change significantly as shown by previous simulations 
(\citetalias{ggc2012,ggc2014,gc2013,ggc2015}), performed using pseudo-Newtonian
potential. We will explore the effects of these dissipations in future studies
using our general relativistic numerical simulation codes and study how the segregation of the 
flow into two components may happen in general relativistic framework.

\section*{Acknowledgements}
DSB acknowledges support via NSF grants NSF-DMS-1361197, NSF-ACI-1533850, NSF-DMS-1622457.
Several simulations were performed on a cluster at UND that is run by the Center for Research Computing. 
Computer support on NSF's XSEDE and Blue Waters computing resources is also acknowledged.









\bsp	
\label{lastpage}
\end{document}